# Quantum Size Effect Exponential Heat Capacity in 4 nm Natural Nickel Nanolattice

Tarachand Patel,[1] Jaiveer Singh,[2] Satya Sanmukharao Samatham,[1,†] Dontula Venkateshwarlu,[1,#] Netram Kaurav,[3] Vedachalaiyer Ganesan[1] and Gunadhor Singh Okram[1,*]

[1] UGC-DAE Consortium for Scientific Research, University Campus, Khandwa Road, Indore 452001, MP, India

[2] Department of Physics, ISLE, IPS Academy, Rajendra Nagar, Indore 452012, MP, India

[3] Department of Physics, Government Holkar Science College, A. B. Road, Indore 452001, MP, India.

**ABSTRACT:** Quantum size effect-induced heat capacity of metal nanoparticles at low temperatures was predicted 79 years ago to be exponential. This, however, has not been reported until date. In defiance, we demonstrate here observation of exponentially decaying heat capacity, *below 45.2 K*, associated with quantum jumps, exceptionally in 4 nm naturally assembled hexagonal closed packed (hcp) lattice of nickel nanoparticles; high magnetic fields have negligible effect on these features. Magnetic susceptibilities in contrast reveal evolution of quantum size effects with decrease in particle size. They exhibit sharp rise below about 30 K and vestiges of saturations below 5 K. The former is explained by Curie-like characteristics of odd electrons while the latter tend towards the orthogonal even-like case. These characteristics, ascribed to the ensembles of Ni nanoparticles, will give a new direction in understanding this crucial thermodynamic phenomenon.

The exponentially decaying heat capacity $C_P$ at low temperature (T→0) of metal nanoparticles for equal level spacing ($\delta$) due to quantum size effect (QSE)[1-3] has not yet been observed experimentally except their reduction[4-7] or enhancement[8-14], compared to their bulk counterparts. This null observance, in contrast to quantum size effects observed in facsimile ways in other various physical parameters[15-19], is attributed to the high sensitivities to uncontrollable influences of their matrices or environment[1-3,8] and to natural difficulties in obtaining ideally monodispersed metal nanoparticles (MNPs)[8,20,21]. The difficulty in the experimental observation of exponentially decaying heat capacity of MNPs has been believed to be due to the simple theoretical assumptions[1-3] that cannot meet the real experimentally challenging environments[3,7,8]. They include (i) a single metal particle not able to meet the thermodynamic nature of heat capacity[3,8], (ii) size effect limitation d ≥ 10 nm with T < 0.1 K to avoid surface effects[8]; (iii) matrix disturbance on metal particles' properties[18,20,22] and (iv) smoothening of level spacings in an ensemble of practically monodispersed particles[3,7,8]. The typical samples used have been compacted gas-/ thermally-evaporated Pd[9,10], V[11,12] nanoparticles (NPs), polyol-method[13] and thermal decomposition[14] prepared Ni NPs, in which at least some surface oxidation naturally occurs[13,20] in addition to the interactions with surfactants/ matrices[20-22].

Stewart[4], for example, reported a reduced heat capacity of Pt NPs embedded in $SiO_2$ at low temperatures and argued it to be a manifestation of QSE. However, majority of enhancement was attributed to the matrix, say $SiO_2$ since Pt and $SiO_2$ were cosputtered. The problem was that he subtracted the bulk values of heat capacity of Pt and vitreous silica from nanoparticle data with the disadvantage of large error involved especially from the sputtered $SiO_2$ by taking the values of vitreous silica directly. Schmidt et al.[5] reported a negative microcanonical heat capacity near the solid to liquid transition on cluster of 147 sodium atoms produced in a gas aggregation source. Volokitin et al.[7] based on the presence of spurious magnetic impurities especially in Pd[5], on the other hand reported the odd-even electron spin states predicted for QSE with the significant influence of the magnetic field on the magnetic susceptibility and specific heat below 0.2 K of shellular Pd clusters but not in colloidal Pd. Moreover, exponentially decaying $C_P$ was ruled out with no quantum jumps. In contrast, in all the other cases such as Al, V, In, Sn, Pt and Pb, enhanced $C_P$ only have been reported and the same have been attributed to either QSEs of the vibrational spectrum of nanoparticles, or an increased electronic heat capacity most likely associated with the surface states of the nanoparticles[8]. Thus, the environment of each system of MNPs in earli-er works finally appear to turn out to be contrary to that required theoretically on quantum size effect of exponentially decaying C_P, leading to null result.

In contrast, we show here the definitively exponentially decaying heat capacity, *below 45.2 K*, associated with quantum jumps in about 4 nm nickel nanoparticles, assembled naturally into hexagonal closed packed (hcp) lattice; yet, magnetic susceptibilities in contrast reveal evolution of QSEs with decrease in particle size. Nickel nanoparticles we studied here were their compacted pellets of average particle sizes in the range ~ 4 - 10 nm prepared using trioctylphosphine (TOP) and oleylamine as surfactants and nickel acetylacetonate as precursor of Ni nanoparticles.

**RESULTS AND DISCUSION**

SAMPLE CHARACTERIZATIONS

Transmission electron microscopy (TEM) images showed monodispersed nature of nanoparticles. Figure 1 (a) shows a typical image of 4 nm sample, which was confirmed from the particle size distribution curve (Figure 1, inset, right bottom panel). Similarly, sizes we studied for other samples thus were 6.0 nm and 10.1 nm. Selected area electron diffraction (SAED) of Ni atoms in 4 nm sample showed only one ring (Figure 1, inset, left bottom panel). The particle sizes determined from Scherrer formula were 1.12 nm, 5.1 nm and 10.1 nm, respectively, with an error of nearly ± 0.01 nm each. Small angle X-ray scattering, SAXS[20] (not shown here) and wide angle X-ray diffraction, XRD (Figure 1(b)) data analysis have identified the nanolattice and atomic lattice structure formations, respectively, of these nanoparticles. Several distinct low angle peaks observed in SAXS data were found from systematic calulations to be due to hcp lattice planes formed by the monodispersed nanoparticles.[20] A single peak observed in wide angle XRD (Figure 1 (b)) matches a single ring seen in SAED pattern (Figure 1, inset, left bottom panel) that however is analytically found through Rietveld fitting (Figure 1 (b)) to be due to a mixture of 12% face-centered cubic (fcc) and 88% hcp atomic lattices formed in these nanoparticles. Further, detailed Rietveld refinement of XRD combined with magnetic susceptibilty data and extended X-ray absorption fine structure analyses of the NPs of other sizes revealed (not shown here) that hcp atomic lattice phase gradually reduces at the cost of fcc atomic lattice phase as particle size increases until 6.0 nm nanocrystals beyond which fcc phase only prevails. Note the contrast of these results to the earlier report,[23] wherein, with a polydispersity of 1-8 nm Ni nanoparticles prepared using reflux of mixture of tertiary alcohol, NaH and ayhydrous nickel salts in tetrahydrofuran, there is a separation in the size distributions of hcp and fcc particles: hcp particles occur at size well below a diameter limit of 3-4 nm while fcc particles occur close this limit or above. This difference is correlated to the variation in the preparation procedure of Ni nanoparticles (or ligand used) in addition to their use of TEM data only. In the present samples, TOP was proven to be responsible for natural nanolattice formation, which was further confirmed from SAED data of the nanoparticles (not shown here) and low value of zeta potential that showed agglomeration of nanoparticles[20]. X-ray photoelectron spectroscopy (not shown here) of Ni further corroborated XRD result of formation of Ni metal with an additional indication of QSE while that of P suggests partial substitution of Ni atoms in its lattice and presence of TOP on the surface of particles. Existence of TOP, oleylamine and acetyleacetonate, by-product of sample preparations, was further confirmed from fourier transformed infra-red spectral analysis. Details of these sample preparations and characterizations were reported in references 20, 21.

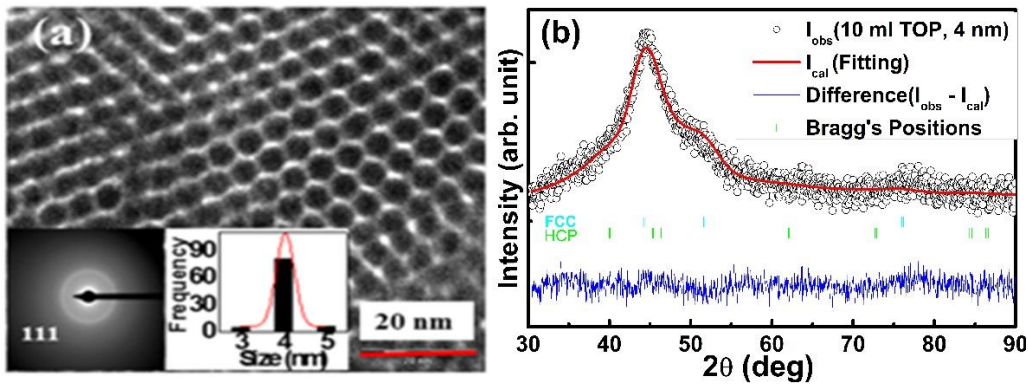

**Figure 1** Characterization of sample. Typical (a) transmission electron microscopy image and (b) wide angle X-ray diffraction pattern (open circles), with Rietveld fitted curve (continuous red), vertical lines indicating peak positions and difference curve between former two curves, of 10 ml trioctylphosphine i.e. 4 nm sample. Insets: selected area electron diffraction data (left panel) and plot of the particle size distribution (right panel), with the latter showing monodispersed size distribution and particle size 4±0.01 nm.

HEAT CAPACITY STUDY

Figure 2 displays the temperature dependence of the representative heat capacity, $C_P(T)$, of 4 nm, 4 nm (R), 6.0 nm and 10.1 nm Ni nanoparticles and Ni bulk as reference: 4 nm (R) sample served the repetition of the measurements to check for reproducibility of the data of 4 nm sample when it was prepared under similar conditions, and similarly the other samples, to assess evolution of $C_P$. Clearly, $C_p$ of the NPs is larger than that of the bulk Ni. It is noted that while $C_P$ curves for smaller NPs are very close to each other, especially above 45 K for 4 nm, 4 nm (R) and 6.0 nm samples. However, $C_P$ for 10.1 nm NPs is quite distinct from those of other samples. It tends to, yet well above, that of the bulk. They in general indicate change in the electronic and phonon contribution to the heat capacity. There is, in overall, enhancement of $C_P$ at 120 K as the particle size decreases. The values are at slight random variation with smaller sizes, indicating that the particle sizes are just an average but the actual size could be varied not only that the surface interactions with the surfactants and environment could be there during and after their preparations. They could drastically influence the final physical property under consideration[18,20,22], say heat capacity.

The most dramatic observation is the exponentially decaying heat capacity $C_P$ of 4 nm NPs below about 45 K (Figure 2). It can be fitted as $C_P = 2336 e^{-140(8)/T}$, expected[1,3] for T→0 and for equal level spacing δ i.e. $k_B T \ll \delta$, to the experimental $C_P$ data below 45 K (Figure 2, inset). This, that too setting in at 45 K, is remarkable since the level spacings here, as will be seen below, are unequal. This is so, in spite of the probable (i) splitting of the degeneracy of the energy levels due to the atomic irregularities on the surface and produce an average level spacing δ, and (ii) deviation (above 17 K) from the original odd-even regime (< 1 K) and hence with their possible wash out. The fitted parameter, $\delta_{fit}$ = 140(8) K, is significantly larger than those quantum jumps observed at and below 45 K (Figure 3). This is attributed to (i) the poor fitting ($\chi^2$ = 32) due to the quantum jumps (Figures 2 & 3) and or (ii) the probable existence of a dominant size of less than 2 nm in the background equivalent to $\delta_{fit}$ = 140(8) K, a little less than double the size determined from Scherrer formula[21], but looks consistent[7]. On the contrary, null results are seen in other samples. They not only look to support, but also strengthen further the earlier null results[4-14]. Hence, the exponentially decaying $C_P$ of 4 nm specimen above 17 K could be a chance that attains near to perfection meeting QSE at such a high temperature.

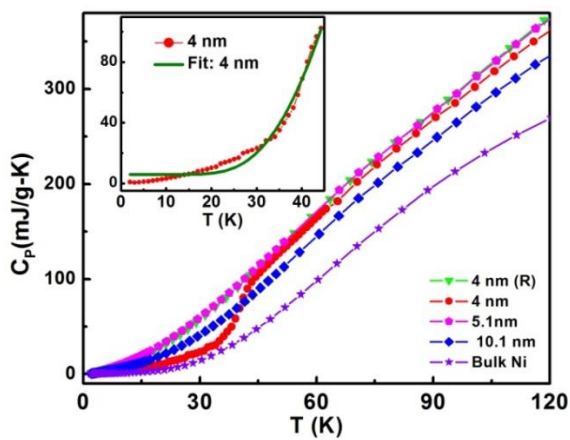

**Figure 2** Heat capacity ($C_P$) data. $C_P$ data of 4 nm, 4 nm (R), 6.0 nm and 10.1 nm natural nickel nanolattice samples along with that of bulk nickel as reference. $C_P$ of 4 nm sample shows an exponential decay characteristic. Inset: Solid (olive) curve is the exponential decay fit $C_P = 2336 e^{-140(8)/T}$ to the experimental $C_P$ data of 4 nm sample (line + symbol).

In order to understand this better, Figure 3 represents excess heat capacity $\Delta C_P$ of 4 nm NPs on the right side ordinate. Its derivative is shown on the left side ordinate. The distinct appearance of the quantum jump features is inferred from these plots. $\Delta C_P$ exhibits abrupt jumps at 45.2 K, 40.2 K, 36.1 K, 32.1 K, 27.1 K, 22.1 K and 17.1 K, respectively. Moreover, the negative excess heat capacity $\Delta C_P$ appeared below 31 K and 34 K is consistent with earlier report on Na cluster[5]. These quantum jumps are seen as peaks in the $d\Delta C_P/dT$ plot and reflect presence of different Kubo energy spacings[2]. They are denoted as $\delta_0$, $\delta_1$, $\delta_2$, $\delta_3$, $\delta_4$ and $\delta_5$. Their respective differences have been defined as $\Delta\delta_1 = \delta_1 - \delta_0$, $\Delta\delta_2 = \delta_2 - \delta_1$, etc., such that $\Delta\delta_1 = 4.1$ K, $\Delta\delta_2 = 4.0$ K, $\Delta\delta_3 = 5.0$ K, $\Delta\delta_4 = 5.0$ K, $\Delta\delta_2 = 5.0$ K, respectively. They are clearly distinct and some of them even have equal values. The various quantum jumps and unequal jump widths ($\Delta\delta$'s) are expected for practical MNPs[2,24,25]. They appear to indicate the discrete nature of set of NPs that possess such level spacings that differ from their smaller or bigger NPs nearly exactly the number of electrons available in each set of these nanoparticles.

This may be understood like this. If bulk Ni atomic density $n=1.818\times10^{23}$/cc is used, the number of electrons N in a 4 nm spherical nanoparticle = nV, with $V = 4\pi r^3/3$ is estimated to be 6092. Then, with $E_F = 11.69$ eV for Ni bulk, Kubo energy spacing, $\delta = 4E_F/3N = 29.7$ K. Conversely, one can readily estimate the particle size from the observed temperature of choice in the range of discussion, and particle size in the range 3.48 nm to 4.8 nm could be statistically possible in this sample of average size 4 nm[2]. The size variation could be reduced if symmetry change of the particles is also considered[3,23]. This is suggestively combined with other ensembles, which have decreasing energy levels and hence corresponding quantum jumps (Figure 4) since particles may have distributed level spacings even for the same estimated size[3,8,25]. They therefore may indicate partially the question of whether there are randomly distributed electronic level spacings over the sample or not[2]. These distinct features are however absent in other Ni nanoparticle samples (Figure S1 in the Supporting Information), but consistent with the restrictive nature of quantum size effects, $k_BT<<\delta$, for all particle sizes studied thus far[2,8].

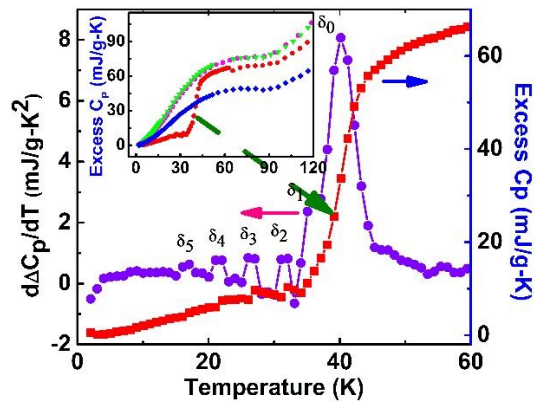

**Figure 3** Quantum jumps in the excess heat capacity $\Delta C_P$ and its derivative $d\Delta C_P/dT$. $\Delta C_P$ of 4 nm NPs (right ordinate) and its derivative (left ordinate). The abrupt quantum jumps in $\Delta C_P$ are seen as peaks in $d\Delta C_P/dT$. They are identified as various level spacings (see text). The inset depicts $\Delta C_P$ of all the four Ni nanoparticle samples to distinguish the anomalous characteristics of 4 nm sample: 4 nm (red rectangle), 4 nm (R) (green triangle), 6.0 nm (pink pentagon) and 10.1 nm (blue vertical rectangle) samples.

The surface atoms interacting with ligands ($A_nT^2$) also contribute to the heat capacity in addition to quantum size effect ($A_0e^{-B/T}$), electronic ($\gamma T$) and phonon ($\beta T^3$) contributions. Here, $\gamma = \pi^2 k_B^2 D(E_F)/3$, $\beta = 12\pi^4 R/5\theta^3$ with $D(E_F)$ is the density of states at $E_F$, R is the universal gas constant and $\theta$ is the Debye temperature. Further, $A_n$, $A_o$ and B are constants with $B = \delta/k_B$. The fitted parameters of the best fits of the expression of these components together to the experimental $C_P(T)$ data are worth mentioning (Table S1 and Figure S2). The value of $\gamma$ increases, but those of $A_n$, $\beta$ and $A_o$ remain nearly constant while those of $\theta$ and B are relatively varied, as the particle size decreases. This is so excluding 4 nm sample which, with distinct QSE in contrast, showed completely different scenario compared to others. Values of $\gamma$, $A_n$ and $\beta$ of 4 nm sample are 1/1.3$^{th}$, 1/20$^{th}$ and 1/10$^{th}$ of those of 4 nm (R) while those of $\theta$, $A_o$ and B are increased to 2-, 600- and 5-fold. This means that due to QSE in 4 nm sample, electronic contribution has decreased marginally while those for surface and phonon contributions have gone down drastically in compensation to significant enhancement in Debye temperature, QSE pre-factor and or QSE effect. These parameters therefore indicate the radical change in the thermodynamic property of 4 nm sample that exhibit QSEs compared to other nickel nanolattice samples. The quantum jump features or the heat capacities, in general, remain nearly intact in high magnetic fields (Figure S3). This is consistent with paired spins in even case[7].

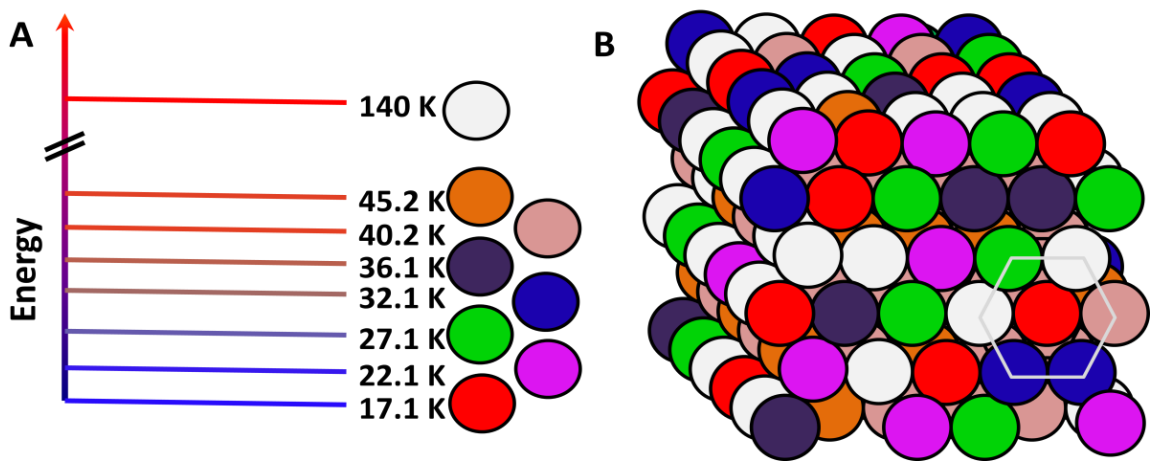

**Figure 4 Energy levels in 4 nm hexagonally closed packed, hcp, nickel particle specimen with faced-centered cubic** *atomic structure* **inside each of them.** A, Energy levels due to various sub-ensembles of nanoparticles represented by white, orange, light red, purple, blue, green, pearl magenta and red color spherical particles responsible for exhibition of exponential heat capacity behavior, quantum jumps at 140 K, 45.2 K, 40.2 K, 36.1 K, 32.1 K, 27.1 K, 22.1 K and 17.1 K, respectively, in the specimen. B, The hcp specimen with different sizes/ types of particles. Particles are shown as of the same size from the point of view of hcp structure formation. However, their response to the heat capacity is different in size based on the transformation properties of the Hamiltonian that is decided by the symmetry, actual size and interaction with their immediate neighborhoods[8].

MAGNETIZATION STUDY

On the other hand, there are systematic setting in, and evolution of, abrupt enhancements in magnetic susceptibility ($\chi$) below 30 K with sign of saturations and then probable decays below about 5 K (Figure 5) as the particle size decreases. These features of susceptibilities and heat capacities observed above 2 K and 17 K, respectively, being at comparatively high temperatures may however have irrelevance of even or odd electrons since they hold at T→0[3]. The former is explained by the internal magnetic fields of about 0.114 T, 0.025 T, 0.013

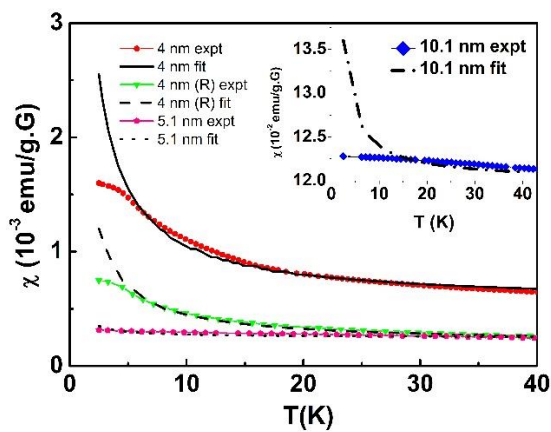

**Figure 5** Magnetic susceptibility $\chi$ data. $\chi$ of 4 nm, 4 nm (R), 6.0 nm and (inset) 10.1 nm natural nickel nanolattices (line + symbol) measured on a mass of nearly 10 mg of piece of a pellet of each sample. The theoretical fits specified in different types of continuous/ dotted curves of the Curie-like behavior of quantum size effect model of odd electron system are shown for verification of the experimental data.

T and 0.558 T at 3 K, respectively of these nanolattices (Table S2). The Curie-like characteristics[3] for odd electrons explain these data reasonably well (Figure 5) rather than an orthogonal ensemble with an even number of electrons (Figure S4; cf. ref. 24). They are in marked contrast to those observed in Pd5 clusters below 0.2 K at high field wherein $C_P$ data were reported to agree with $T^2$-dependence predicted for QSE but ruled out the exponential decay of $C_P$ at T→0[7]. The presently observed characteristics are attributed to the natural nanolattice of colloidal nickel particles, which are monodispersed such that the influence of the ligands are limited to surface[25], and to interparticle cohesion assembling the particles into hcp crystal structure[20] without disturbing the distinct characteristics of each sub-ensemble of Ni nanoparticles. Hence, their characteristics of fulfillment of near to what is required theoretically that too most distinctly in one sample only yet at much higher temperatures, one to two orders of magnitude, clearly demonstrating this very fundamental aspect of single particle thermodynamic properties[2], are outstanding.

CONCLUSIONS

Heat capacity and magnetization measurements were carried out on the compacted naturally formed monodispersed nickel hcp nanoparticle lattices, i.e. nanolattices, of particle sizes 4 nm, 4 nm (R), 6.0 nm and 10.1 nm; 4 nm (R) was a repetition of 4 nm sample for reproducibility. Quantum size effect-induced exponentially decaying heat capacity, *below 45.2 K*, associated with quantum jumps, was observed exceptionally in 4 nm, naturally assembled hcp lattice of, nickel nanoparticles. These features as well as heat capacity data of other samples were found to have negligible effect of high magnetic fields. On the contrary, the magnetic susceptibilities reveal evolution of quantum size effects with decrease in particle size. They exhibit sharp rise below about 30 K and vestiges of saturations below 5 K. The former can be explained by Curie-like characteristics of odd electrons while the latter tend towards the orthogonal even-like case. The exponential quantum size effect characteristic that is ascribed to the ensembles of Ni nanoparticles, exemplifying the nearly theoretical one particle nature of heat capacity of metal nanoparticles, is an extraordinary observation, pending to reveal experimentally since its prediction. This will provide a new vista on the peculiarity on this thermodynamic property.

**SAMPLE PREPARATION**

A thermal decomposition method was used to synthesize the nanoparticles. Typically, 10 ml (i.e. 22.4 mM) of preheated (215 °C) trioctylphosphine (TOP, 90% Aldrich) was added in the already degassed (at 100 °C for 30 min) solution of 1.02 g Ni(acac)$_2$ (95% Aldrich) and 8 ml oleylamine (OAm, 70% Aldrich). The resulting solution was further heated at 220°C for 2 h under argon atmosphere. This gave rise to black precipitate due to the formation of nickel nanoparticles. Solution was then cooled to 27 °C, and centrifuged by adding ethanol (99.9% Jiangsu Huaxi) to extract and wash the nanoparticles. Washing was performed four times. Similar procedures were followed for 1 ml, 5 ml and 10 ml (repeated sample, R) of TOP at fixed OAm (8 ml). The particles were dried at 60 °C and used directly for characterizations. The particle sizes mentioned here were estimated from transmission electron microscopy. Details of sample preparation and characterization can be found elsewhere[20,21]. It may be noted here that the Ni nanoparticles used here are not only monodispersed but also naturally self-assembled themselves into their own lattice i.e. lattice of nanoparticles or nanolattice, in addition to the atomic lattice, whenever trioctylphosphine is one of the surfactants or is the only surfactant[20,21].

METHODS

The Bruker D8 Advance X-ray diffractometer with Cu K$\alpha$ radiation (0.154 nm) in the angle range 30-90° was used for laboratory method of XRD measurements of the samples in powder form; the X-rays were detected using a fast counting detector based on silicon strip technology (Bruker LynxEye detector). Nanoparticle images and selected area electron diffraction (SAED) were recorded using transmission electron microscopy (TECHNAI-20-G$^2$) by drop-casting the well-sonicated solution of a few milligrams of nanoparticles dispersed in about 5 ml ethanol on carbon-coated TEM grids. TEM was operated at 120 keV. At least 200 particles were analyzed per sample to obtain a representative size distribution. Heat capacity was measured on nearly 10 mg in mass of a piece of pellet of each sample using relaxation calorimetric method (Quantum Design). Each measurement was analyzed using a two-$\tau$ model to accurately simulate the effect of the heat flow between the microcalorimeter platform and the sample ($\tau_2$) as well as the heat flow between the platform and puck stage ($\tau_1$). Magnetic susceptibility was measured using superconducting quantum interference device – vibrating sample magnetometer (SQUID-VSM, Quantum Design) at 50 Gauss in the temperature range 300 K down to 2 K.

## ASSOCIATED CONTENT

**Supporting Information**. More detailed analysis with Figures S1-S4 and Tables S1-S2 on the various theoretical fittings and parameters obtained from the experimental heat capacity and magnetization data.

## AUTHOR INFORMATION

### Corresponding Author

*Author to whom correspondence should be addressed. E-mail: okramgs@gmail.com, okram@csr.res.in

### Present Addresses

†Present affiliation: Magnetic Materials Laboratory, Department of Physics, Indian Institute of Technology Bombay, Powai, Mumbai 400076, Maharashtra, India. Email: sssrao@phy.iitb.ac.in.

# Present affiliation: Indira Gandhi Centre for Atomic Research, Kalpakkam, Tamil Nadu, India.

### Author Contributions

GSO planned the experiments, did the detailed data analysis, drafted and wrote the paper while JS made the samples as GSO proposed, TC, NK and JS assisted in fitting and analyzing some of the data, and VG measured the heat capacity data along with SSS and DV. All authors have read the manuscript and approve for publication.

### Funding Sources

UGC-DAE Consortium for Scientific Research, Indore, India.

Competing financial interests
Authors do not have any Competing financial interests.

## ACKNOWLEDGMENT

Authors gratefully acknowledge N. P. Lalla, M. Gupta and R. J. Choudhary of UGC-DAE CSR, Indore for the TEM, XRD and magnetization data, respectively.

# Supplementary Information

# Quantum Size Effect Exponential Heat Capacity in 4 nm Natural Nickel Nanolattice

Tarachand Patel,[1] Jaiveer Singh,[2] Satya Sanmukharao Samatham,[1,†] Dontula Venkateshwarlu,[1,#] Netram Kaurav,[3] Vedachalaiyer Ganesan[1] and Gunadhor Singh Okram[1,*]

[1] UGC-DAE Consortium for Scientific Research, University Campus, Khandwa Road, Indore 452001, MP, India
[2] Department of Physics, ISLE, IPS Academy, Rajendra Nagar, Indore 452012, MP, India
[3] Department of Physics, Government Holkar Science College, A. B. Road, Indore 452001, MP, India.

**Supplementary Materials:**

Figures S1-S4

Tables S1-S2

1. **Heat capacity data**

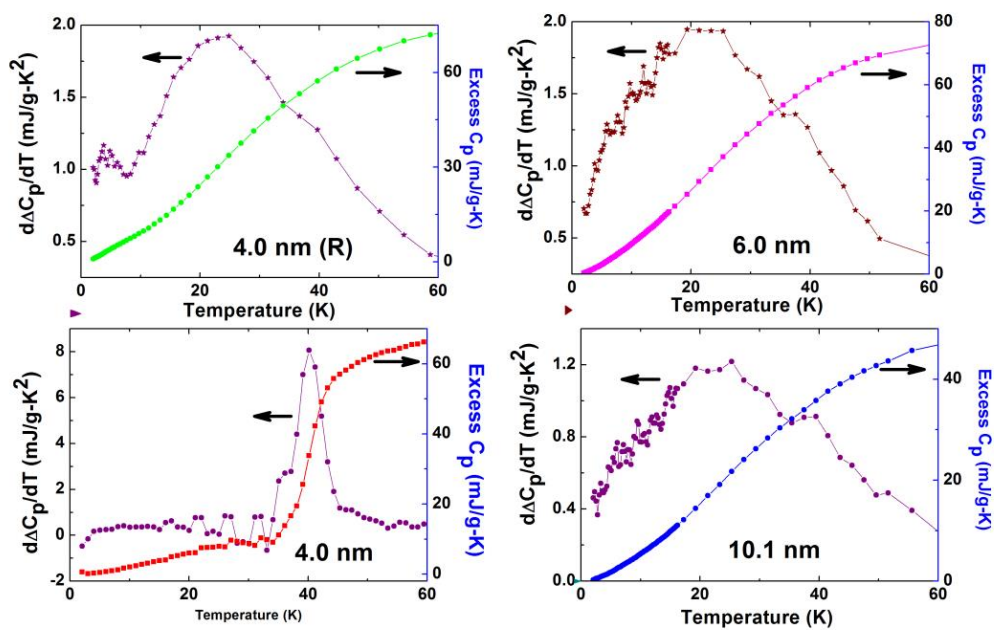

**Figure S1** Quantum jumps in the excess heat capacity $\Delta C_P$ and its derivative $d\Delta C_P/dT$. $\Delta C_P$ of 4 nm NPs (right ordinate) and its derivative (left ordinate). The abrupt quantum jumps in $\Delta C_P$ are seen as peaks in $d\Delta C_P/dT$. They are identified as various level spacings (see text). Similar plots are shown for 4 nm (R), 6.0 nm and 10.1 nm samples. Note their features of $d\Delta C_P/dT$ without any clear peaks even the one near 5 K in 4 nm (R) sample. Only 4 nm sample exhibits clear quantum size effect.

**Table S1** Fitted parameters of experimental heat capacity data considering various contributions (see text.)

| Sample | γ (mJ/g-K$^2$) | $A_n$ (mJ/g-K$^3$) | β (mJ/g-K$^4$) | θ (K) | $A_0$ (mJ/g-K) | B (K) | χ$^2$ |
|---|---|---|---|---|---|---|---|
| 10.1 nm | 0.21 | 0.035 | 0.0001 | 268.8 | 10 | 60 | 4.39 |
| 6.0 nm | 0.45 | 0.04 | 0.0001 | 268.8 | 10 | 20 | 3.71 |
| 4 nm (R) | 0.51 | 0.04 | 0.00011 | 260 | 10 | 40 | 3.85 |
| 4 nm | 0.4 | 0.002 | 0.00001 | 579 | 6000 | 200 | 22.72 |
| Bulk Ni | 0.021 | | 0.00052 | 253.6 | | | 5.8 |

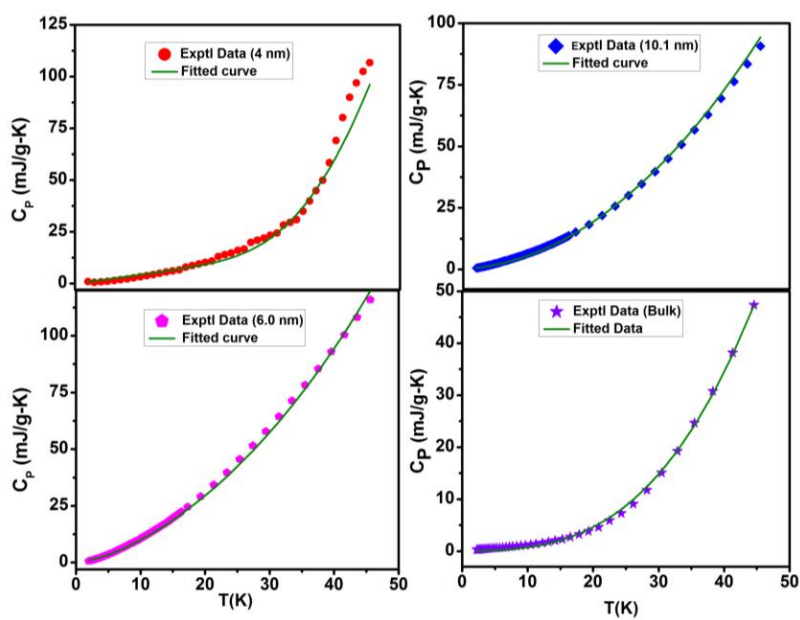

**Figure S2** Theoretical fits (line) of heat capacity of 4 nm, 4 nm (R), 6.0 nm, 10.1 nm nickel nanolattices and bulk of nickel (symbol) considering various contributions (see text).

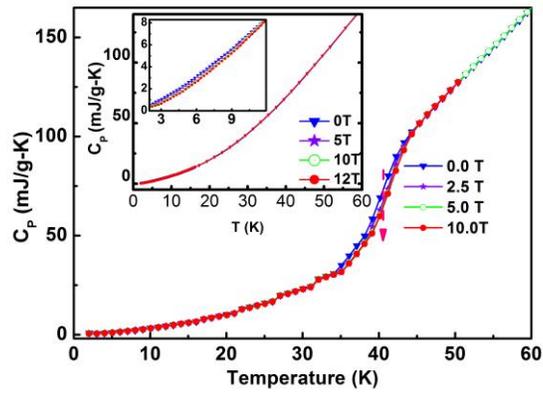

**Figure S3** Influence of high magnetic field (H) on heat capacity. Shown here is for 4 nm sample: arrow indicates the increasing applied field. Inset: $C_P$ for 10.1 nm sample and its magnified portion below 12 K (sub-inset). Note the very feeble influence of H. This is attributed to possible even electron particles for low temperature (T). Since it is so at high temperature, this possibility may not however be applicable.

**Magnetization data**

**Table S2** Blocking temperature ($T_B$), saturation magnetization ($M_s$) and coercive field ($H_c$) of MNPs.

| Samples | $T_B$ (K) | $M_s$ (emu/g) | | $H_c$ (Oe) | |
|---|---|---|---|---|---|
| | | 3 K | 300 K | 3 K | 300 K |
| 10 nm | 167 | 26.9 | 23.5 | 234.3 | 35.0 |
| 6.0 nm | 2.3 129 | 0.6 | 0.22 | 160.5 | 30.9 |
| 4 nm (R) | 4.5 | 1.2 | 0.08 | 190 | -- |
| 4 nm | 4.9 | 5.5 | 0.80 | 203 | -- |

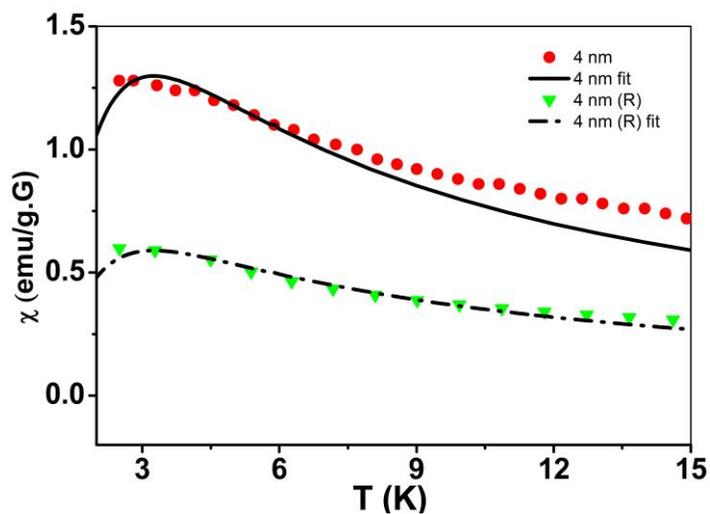

**Figure S4** Magnetic susceptibility ($\chi$) of 4 nm (circle) and 4 nm (R) (triangle) samples, and fits (dotted/ continuous line) of an orthogonal ensemble with an even number of electrons in the temperature range 2 to 15 K. Fittings are not so good. Situations were even worse in the $\chi$ data fits of 6.0 nm and 10.1 nm samples giving rise to the unphysical fit parameters. The origin for not able to fit the data well is attributed to high temperature (T > 2 K) regime where irrelevance of odd or even electrons prevails.